# Immercity: a curation content application in Virtual and Augmented reality

Jean-Daniel Taupiac,[1,2][0000-0003-1018-933x] Nancy Rodriguez[0000-1111-2222-3333] and Olivier Strauss[2]

[1] Capgemini Technology Services, Bayonne, France
[2] LIRMM ï University of Montpellier, Montpellier, France
{jean-daniel.taupiac, nancy.rodriguez, olivier.strauss}@lirmm.fr

**Abstract.** When working with emergent and appealing technologies as Virtual Reality, Mixed Reality and Augmented Reality, the issue of definitions appear very often. Indeed, our experience with various publics allows us to notice that technology definitions pose ambiguity and representation problems for informed as well as novice users.
In this paper we present Immercity, a content curation system designed in the context of a collaboration between the University of Montpellier and CapGemini, to deliver a technology watch. It is also used as a testbed for our experiences with Virtual, Mixed and Augmented reality to explore new interaction techniques and devices, artificial intelligence integration, visual affordances, performance, etc. But another, very interesting goal appeared: use Immercity to communicate about Virtual, Mixed and Augmented Reality by using them as a support.

## 1 Introduction

The Montpellier Laboratory of Informatics, Robotics and Microelectronics (LIRMM) and Capgemini Company work together on a PhD project which focuses on the study Virtual, Augmented and Mixed Reality environments for training. In this context, one of the first difficulties encountered was to communicate about these technologies definitions. Indeed, it soon became clear that definitions posed understanding, ambiguity and representation problems as well for informed than novice interlocutors.

It seemed interesting, in the scope of this project, to disseminate a technology watch destined for colleagues, partners and customers. Today, content curation is principally made on specialized websites and blogs, social or business-social networks as Twitter [1] or LinkedIn [2] as well on dedicated platforms and tools as Scoop-it [3] or Paper.li [4].

Thereby, by seeking to find a way to diffuse the technology watch in an original manner, we conceived a multi-technological application which manages the idea of communicating on these technologies by their use. Our system, **Immercity,** aims to centralize information from the technology watch within a unique 3D representation, a city. By interacting with each building, the user has access to a particular kind information (news, scientific papers, videos), the city becoming in this way a metaphor of a blog or website main menu. The city 3D representation is avalaible in the Virtual and Augmented Reality clients by using a web browser or by installing an application on a smartphone. This multi-technological aspect supports the communication main objective with a secondary one: illustrate differences between these technologies.

In this paper, we introduce our work in progress regarding the development of Immercity. We introduce the general principle of the application and we detail the prototypes as well as our work concerning visual cues for selection and navigation. Finally, we present the first experiment we have run and discuss the results obtained and the improvement perspectives they bring.

## 2 Background

In the context of this work, it seems important to remind that an Augmented Reality system respects three essential rules [5]:

- Combine real and virtual
- Interact in real time by an interactive way
- Be recorded in 3 dimensions

Augmented Reality aims to complete the user perception by adding virtual information. Mixed Reality refers to a continuum connecting physical world to virtual world, including therefore two aspects [6]: Augmented Reality and Augmented Virtuality, which consist of enriching virtual world by adding real elements.

[7] propose a technical definition of Virtual Reality as *a scientific and technical field exploiting information technologies and material interfaces in order to simulate in a virtual world the performance of 3D entities which are in interaction in real time, amongst themselves and with one or several users in a pseudo-natural immersion through sensorimotor channels.*

By relieving previous works [8], [9] extends this definition by determining a fundamental principle, the ñperception, cognition, actionò loop: in any Virtual Reality application, the user is in immersion and interaction with a virtual environment. He perceives, decides and acts in this environment.

As far as we know, the idea of communicate by these technologies in this specific use case has not been exploited yet. However, some interesting initiatives designed to communicate on these technologies by example exists, like the VENTURI [10] system. One of the studies of this project aims to understand what users think about a new concept of AR gaming by using several prototypes.

Curation means to collect and organize resources with value added by an expert in order to lead to greater understanding and insight of information. The Information Visualization domain provides some examples of collection presentation in Virtual Reality. In their CyberNet project, Russo Dos Santos and al. [11] implemented a city metaphor for NFS data visualization. The advantages highlighted by the authors are that a city implies a natural hierarchy (districts, blocks, streets, and buildings) which allows interesting possibilities for hierarchical visualization.

Sparacino and al. [12] designed City of News, an immersive and interactive web browser. It is a dynamically growing urban landscape where information, i.e. URLôs content, is mapped on skyscraperôs frontages. The city is organized in urban quarters and each quarter is linked to a specific thematic. It also evolves and grows organically through exploration: by following a link, the user causes a new city-element creation. Sparacino and al. [13] have early developed Hyperplex, an environment for browsing digital movies. They are structured within a multi-dimensional virtual inhabited building. Each room of the building is associated with specific topics. When a movie content is selected, the background change dynamically by transforming itself according to the content.

In Immercity application, the city is a collection of collections: the buildings are not supposed to represent an information but a collection of informations organized by type.

# 3 The Immercity application

## 3.1 Use cases

As stated before, Immercity aims to be a multi-technological application. Indeed, the targeted users range from Capgemini co-workers to the company partners and customers. Furthermore, the system has to take into account that it can be used in a fixed environment on the workplace as in a mobile context for use on exhibition or conferences.

Following this aspects, several uses cases has been defined for Immercity:

1. On a web browser, the application would allow to display a 3D perspective view of the city and to access the information.
2. In Virtual Reality, the user, equipped with a Cardboard, could explore the city in order to access the information.
3. In Augmented Reality, it would be possible to bring out a 3D model of the city from a visual tag representing its 2D plan. This visual marker, limited in size, can be placed on the back of a business card. The user could then use this tag in order to download the application, and to see the city emerge on 3D.
4. In Mixed Reality, the user visualizes the city which would appear over a planar surface of his environment, or over the same visual tag used in Augmented Reality. He would use his hands to interact with the different objects of the scene. *At present time the Mixed Reality development has not yet started.*

## 3.2 The 3D city representation

By interacting with each building, the user would access to the different information available in the technology watch. We choose to link six key buildings to the differ-ent kind of information, by their semantics and common use (**Erreur ! Source du renvoi introuvable.**):

- The School, in which we find information about the different definitions and concepts.
- The Kiosk, presenting the last news on Virtual, Augmented and Mixed technologies.
- The Supermarket, containing a catalog of existing devices, and supporting comparison based on their characteristics.
- The Library, offering access to scientific papers
- The Cinema, giving access to videos and demonstrations.
- The house, allowing to access userôs preferences and bookmarks.

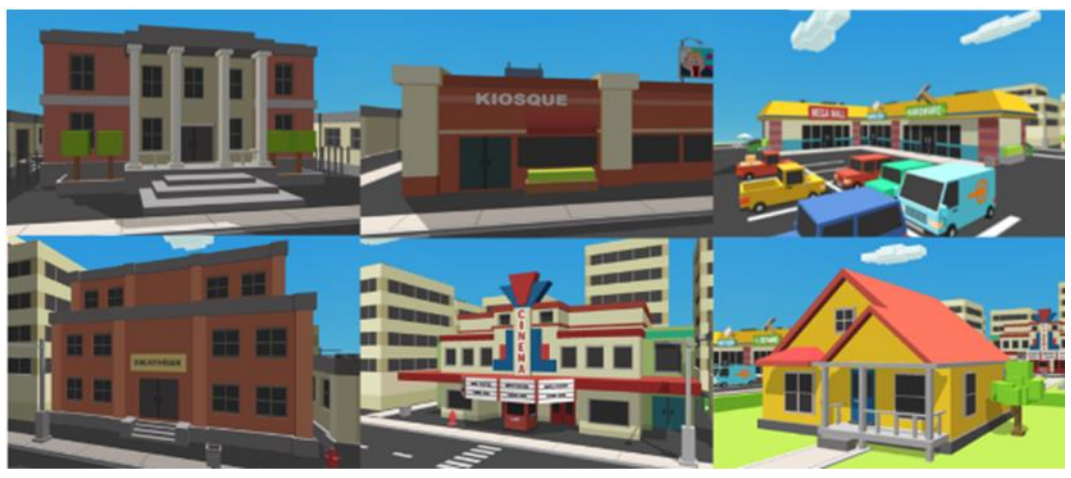

**Fig. 1.** Immercity key buildings

We choose to place key buildings near each other to increase their visibility and ease exploration of content. Furthermore, it was necessary to find a way to highlight key buildings in order to ease their identification by the user. We discuss these visual cues in the section 3.4.

Prototypes for our three first use cases (cf 3.1) have been realized via Unity3D[14], on an eight weeks sprint[15]. The aim was to confirm the technical feasibility of this kind of application. For the city modeling, we choose to keep a low realism level and enhance visual appealing by using a cartoon graphic style.

### 3.3 Web browser prototype

The web browser prototype allows to display a perspective 3D view of the city in a web browser by WebGL (Figure 3) and to select a building by a mouse click. A fixed camera has been positioned in order to visualize the set of key buildings.

To easily identify key buildings, they are surrounded with a halo which changes its color when the mouse hovers over the building. A tooltip with the name of the building is displayed in the same time.

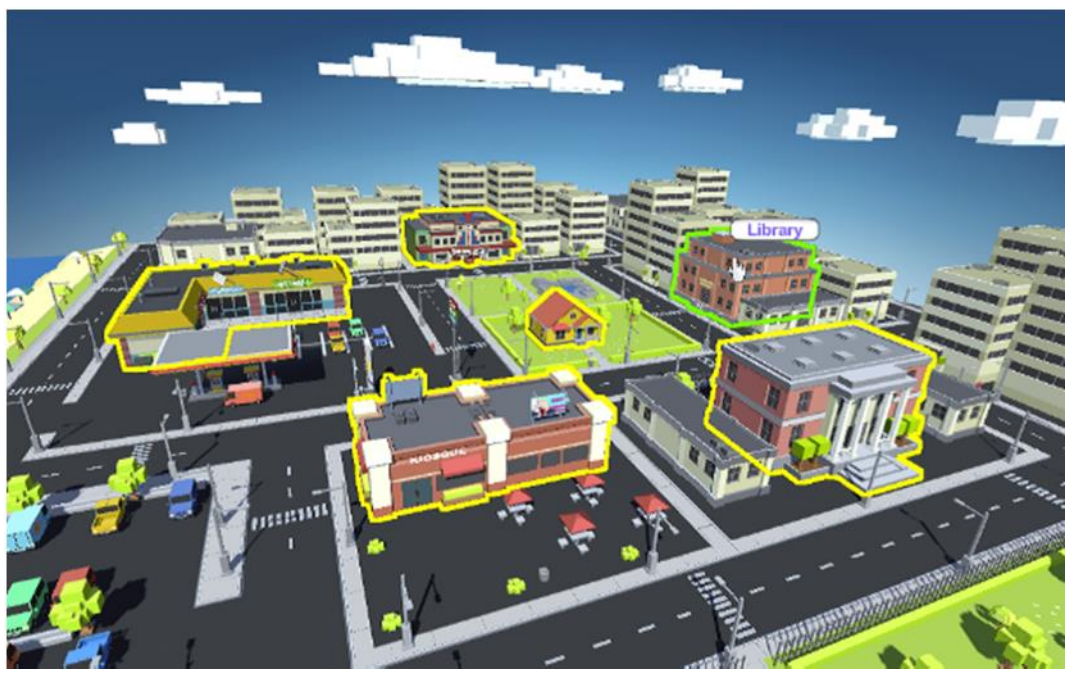

**Fig. 2.** Web browser view

### 3.4 Virtual Reality prototype

The Virtual Reality prototype offers an immersive visualization to users equipped with a cardboard device. It used the Google VR SDK[16] for Unity3D to define the stereoscopic view.

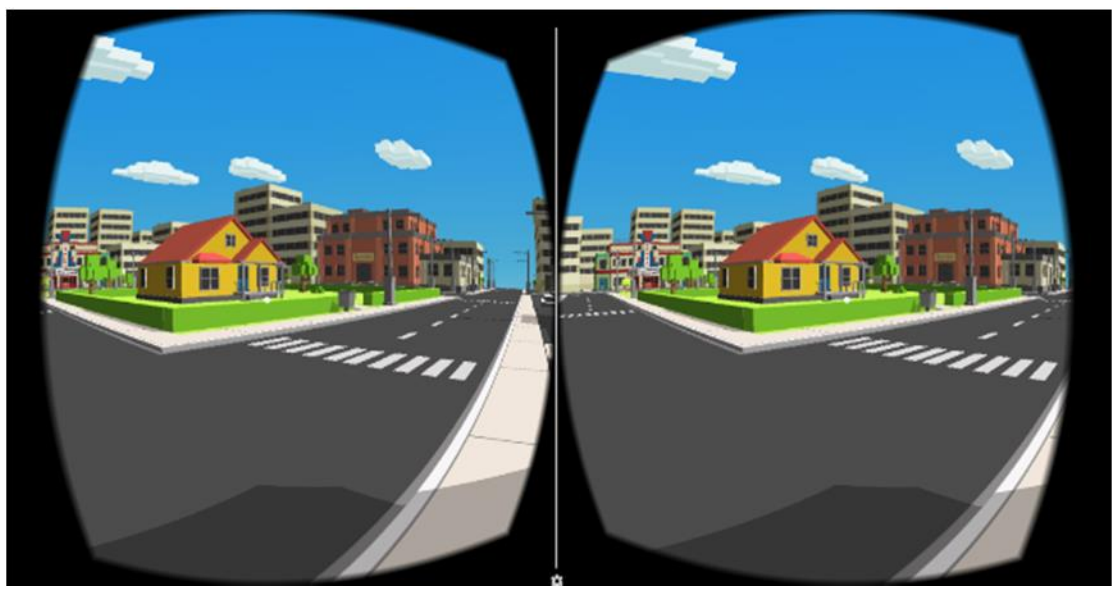

**Fig. 3.** Virtual Reality prototype

To allow navigation while being compatible with different kinds of cardboard devices, we use a visual pointing method. This avoids using cardboard specific physical buttons. The cursor supporting selection and navigation included in the Google VR SDK consists in a small reticle which became wider when it is over an interactive object (Figure 5).

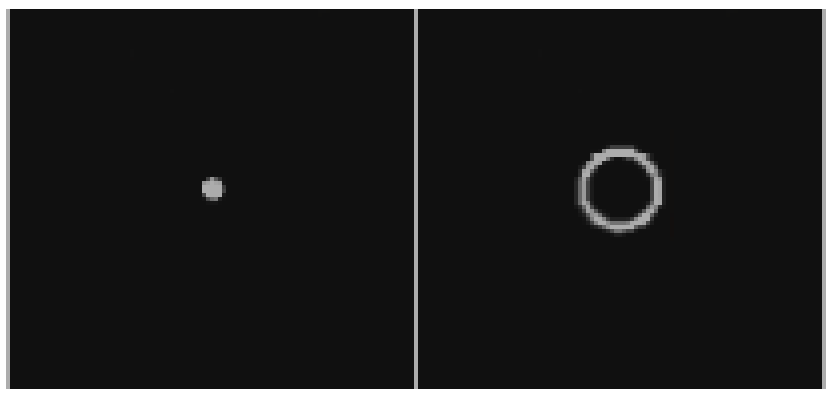

**Fig. 4.** Google VR SDK cursor

This cursor shows however some limits. During the unit tests, it seemed important to have an indicator of the time needed to point an object before its activation. We suppose that novice users could not directly understand that they have to point for a time in order to be transported to the building. We modified the cursor in order to add a second circle, indicating the loading time necessary for triggering the action (Figure 5).

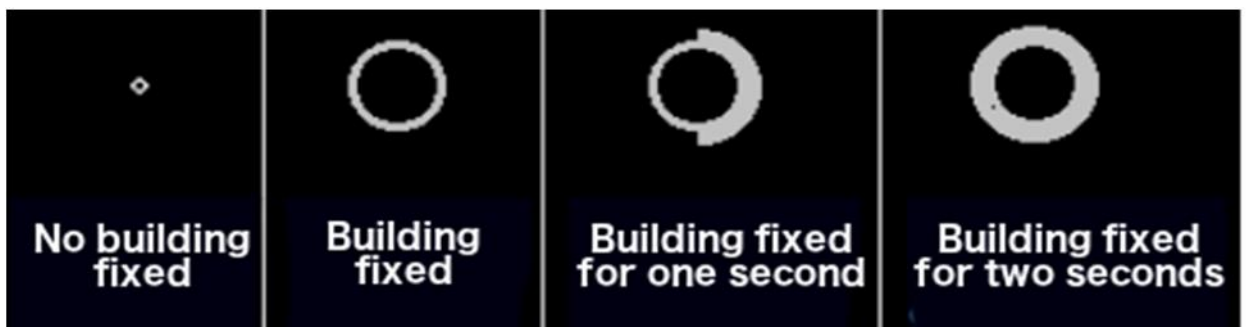

**Fig. 5.** Immercity cursor

As indicated above, the user is conducted to the building she has selected but she has the possibility of change her direction. The moving method was implemented through the A* Pathfinding API [17]. This library allows to define a graph of allowed footpath in the Unity3D scene (Figure 6) and then to move from a point to another. This allows us to respect some rules for example to use the crosswalks. The path is calculated according to the A* pathfinding algorithm, which find the short path between the position of the user and the chosen destination.

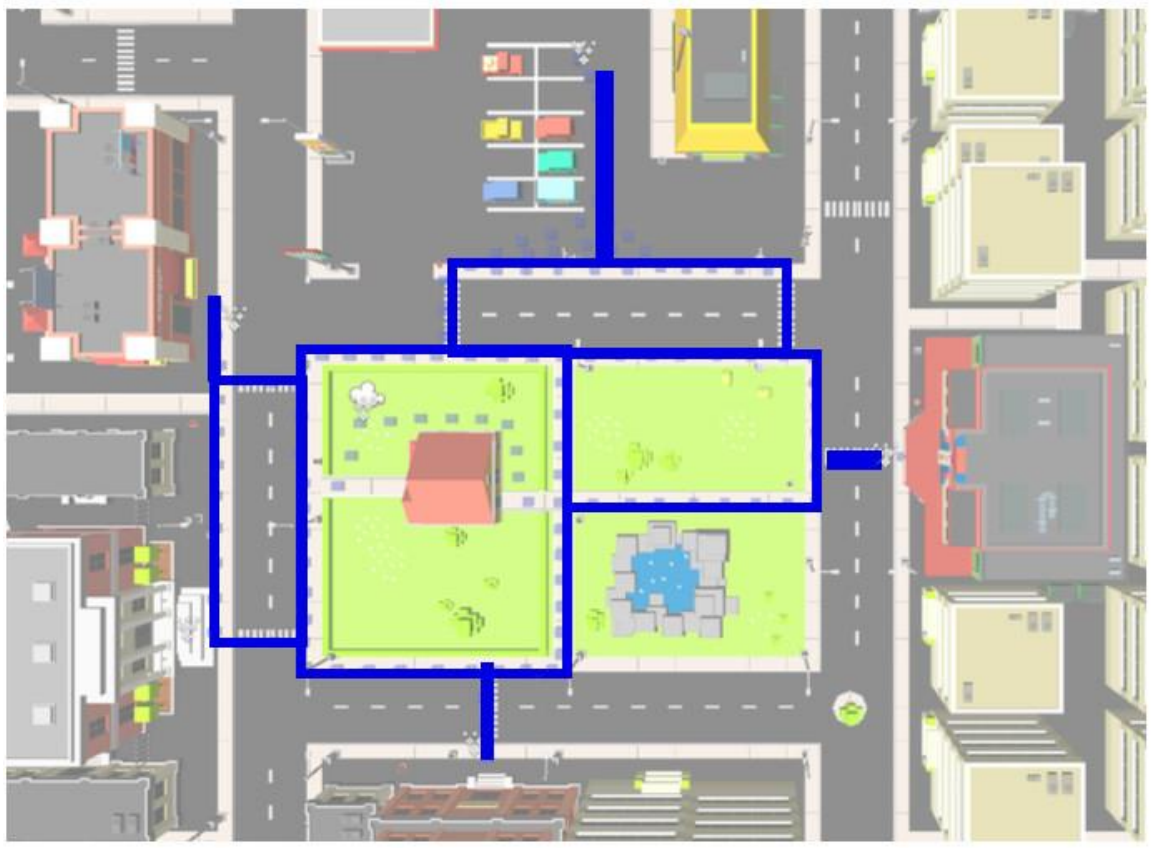

**Fig. 6.** Allowed paths (in blue) definition

As in the web browser prototype, we defined a way to highlight key buildings in order to allow the user to recognize them. It consists in positioning a highlight effect over or in front of each key building (Figure 7).

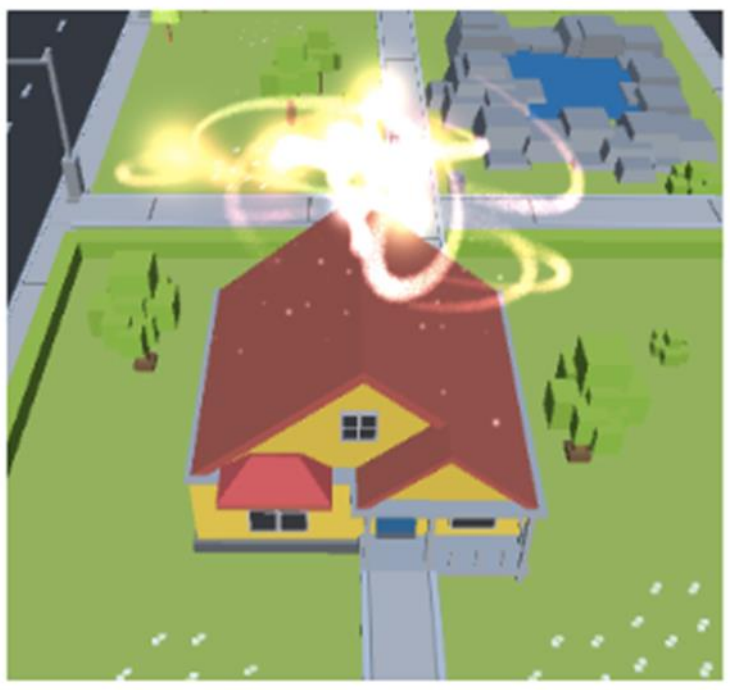

**Fig. 7.** Highlighting key buildings

### 3.5 Auugmented Reality prototype

In the Augmented Reality prototype, the city is displayed on top of a 2D marker. It has been implemented using Vuforia SDK [18], a library allowing to rapidly creating Augmented Reality applications on Unity3D.

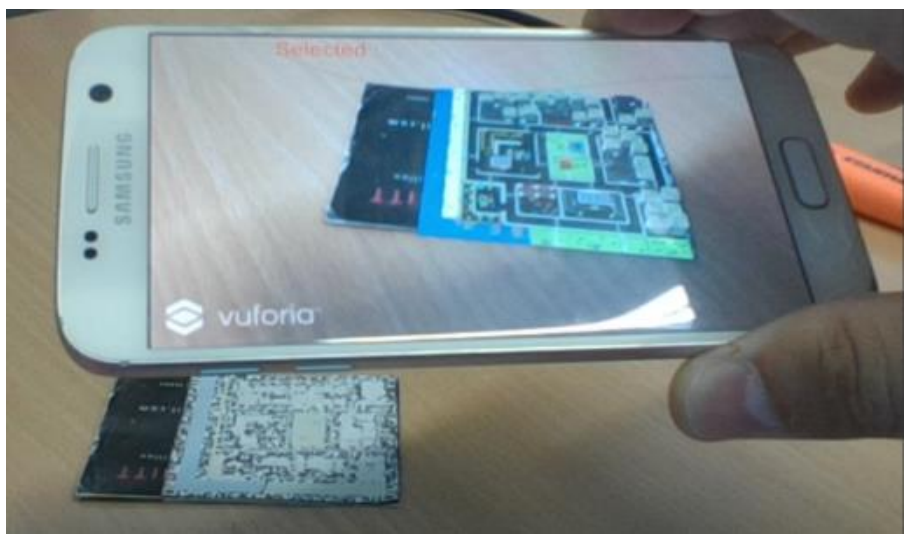

**Fig. 7.** Augmented Reality prototype

Users can interact with buildings in two different ways. The first one, which seems more intuitive, involved a simply tap on the building. In the second one, we integrate the same cursor than in the Virtual Reality prototype. Key buildings are highlighted in the same way as in the Virtual Reality prototype.

One of the main problems of this prototype concerns the visual marker quality, which will be placed in the back of a business card. The initial idea was to use the cityôs 2D plan, but this lightly contrasted image has only a few features (detectable points) and obtained a low reliability indication from Vuforia (2/5).

To add more features points we choose to insert a QRCode in the middle of the card. Even if the number of detectable points increased, detection was not robust and the city model was easily lost when the camera explore the peripheral zones of the city where the QRCode is not visible. A satisfying solution was found by drawing the entire marker in black and white and by integrating directly several QRCodes in the city map (Figure 8). This kind of marker has numerous features, uniformly distributed across the image which allows recognition even when the entire marker is not visible. Therefore it obtained a 5/5 reliability indication from Vuforia.

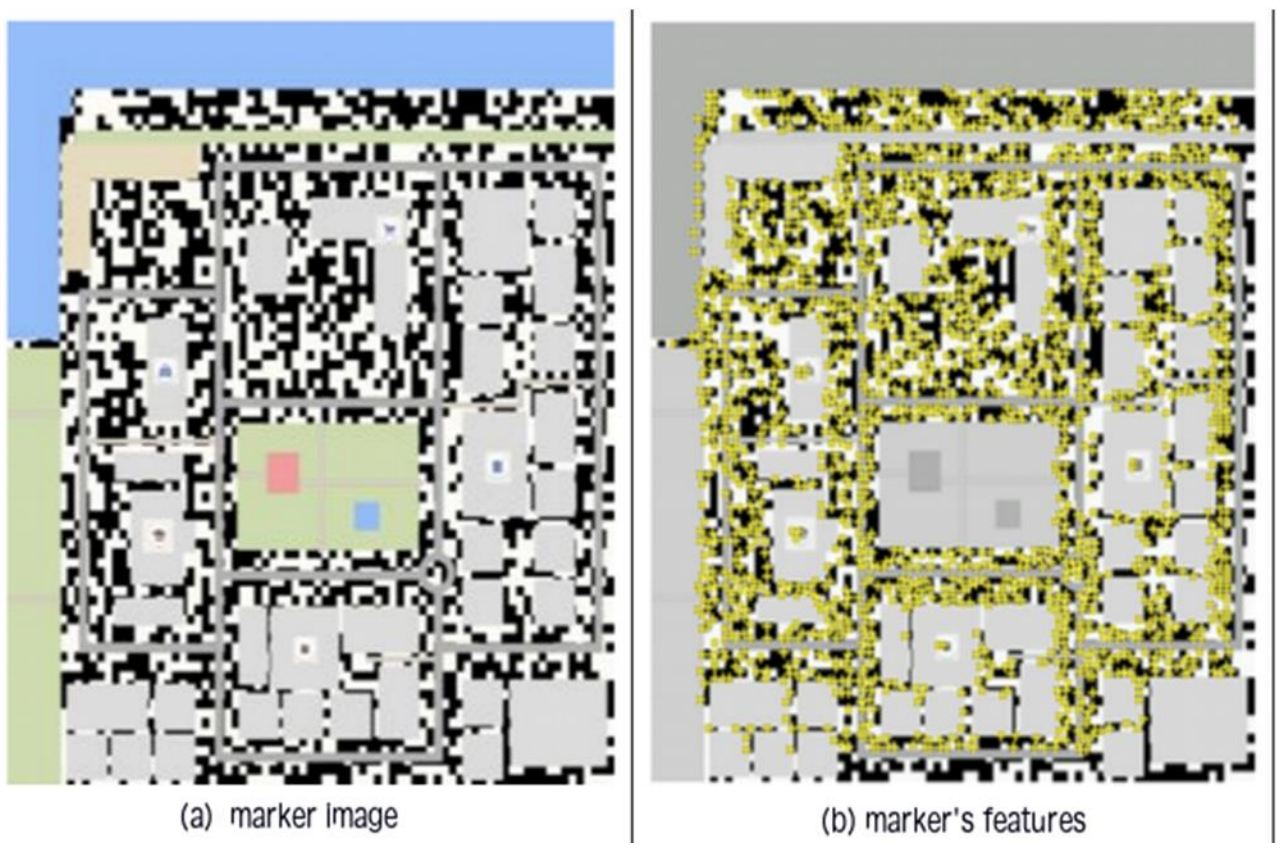

**Fig. 8.** Visual marker for the AR prototype

# 4 EXPERIMENTS AND RESULTS

The goal of our experiment was to analyze the userôs reactions to prototypes and to assess the validity of the visual cues implemented: highlight of key buildings in the Augmented Reality (AR) and Virtual Reality (VR) prototypes, and the visual cursor supporting navigation for the Virtual Reality prototype.

Therefore, we wanted to verify the following hypothesis:

1. In Virtual Reality, users should prefer to use the loading cursor than the simple one included in the Google VR SDK;
2. In Virtual Reality, users should prefer to directly interact with key buildings more than with highlights
3. In Augmented Reality, users should find useful the presence of highlights on key buildings

We didnôt include the Web browser prototype in the experiment, its functionalities being covered for the AR and VR prototypes.

## 4.1 Context and procedure

The experiment took place in the Capgemini office of Bayonne. The panel consisted of 37 persons, mainly Capgemini employees (90%), 70% men. A majority of them (59%) occupy technical posts (Developer or Technical head). Their age varies from 21 to 59 years old, for an average of 36 years old and with an important part of 20-30 years old (48%). Finally, it is interesting to specify that, for most of them, it was their first experience in Virtual Reality (68%) and Augmented Reality (70%).

Experiments were made on a Samsung Galaxy S7 smartphone and, for the VR prototype, using the Samsung Gear VR (v1) headset.

In order to verify our hypothesis, we have defined four test sets for the VR prototype, one for each combination of cursor and highlights:

1. Simple cursor and pointing on highlights
2. Simple cursor and pointing directly on buildings
3. Loading cursor and pointing on highlights
4. Loading cursor and pointing directly on buildings

Each participant tested the two prototypes and one combination of the test sets: 1 and 4, or 2 and 3. For the AR prototype only two test sets were implemented, with highlights enabled or disabled. In order to balance the tests, the proposed test sets were alternated from one subject to another.

Two experimental procedures have been defined. In the first one, the VR prototype integrating one test set was tested for about five minutes. Then we asked the user to cite the key buildings she has been identified. After a brief presentation of key buildings, we asked the user to move towards a specific building. This task was repeated once using a different test set. Finally, the AR prototype was tested. In the second procedure, users tested the AR prototype for about 3 minutes. As before, we asked the user to cite the key buildings she has been identified before a brief presentation of key buildings. Then we asked the user to select a specific building. After activation/deactivation of highlights, the task was repeated. Finally, the VR prototype was presented.

Even if exchanges was directed by a questionnaire (based of the Group Presence Questionnaire [19]), we encouraged subjects to freely discuss during and after the whole procedure in order to collect their impressions and suggestions.

### 4.2 Virtual Reality results

Analysis of results allows us to validate our first hypothesis, it is necessary to provide a loading time indicator. On the 27% of subjects which experienced difficulties in finding the way to move, 70% of them had a test set with the simple cursor. This tendency is confirmed when we analyze the preferences expressed by the subjects (Figure 9). Furthermore, 24% of the subjects express the interest to have a loading time indicator, independently of the cursor type they experienced.

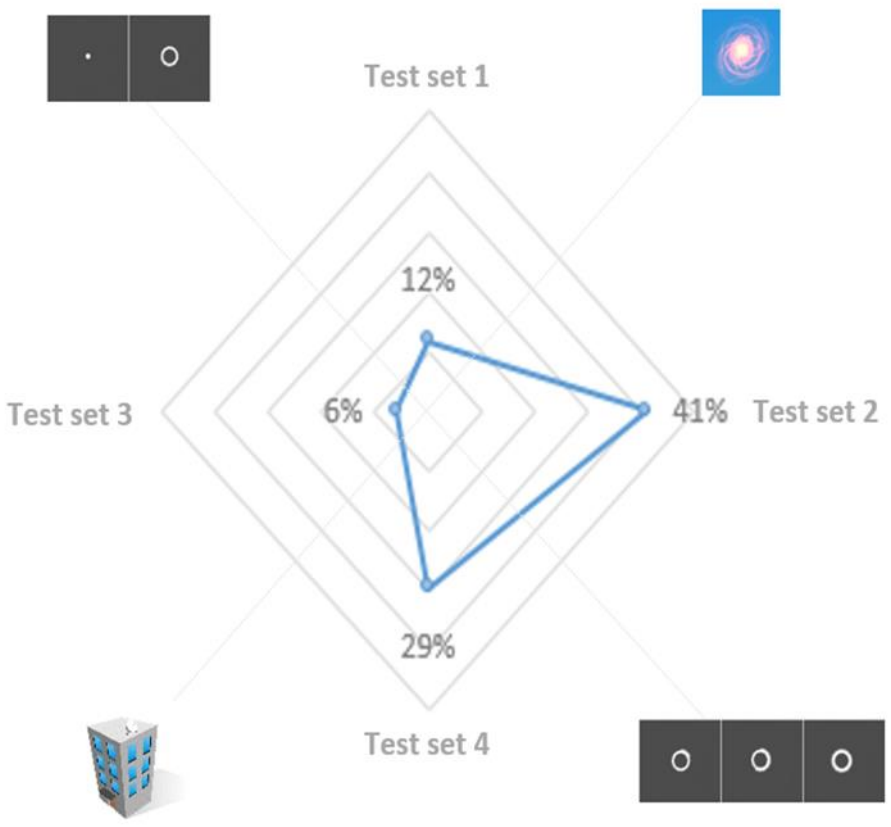

**Fig. 9.** Users preferences in Virtual Reality

Among the subjects having experienced difficulties in finding the way to move, 80% of them had a test set where it was necessary to select the building directly. Nevertheless, we can notice that subject preferences are rather balance on this point. Indeed, some of them questioned highlights on other criteria (aesthetics, practicality, positioning or visual overload). Nevertheless, this result doesnὰt allow us to confirm our second hypothesis.

A large part of users (41%) described the prototypes as ñfunò and ñexcitingò but a substantial part of them was not comfortable with the travel technique proposed. The fact of impose a path has not been appreciated (49%), they want to being teleported (22%), move by themselves (14%), stop the movement (14%) or fly (3%). Future research is needed to address these difficulties.

### 4.3 Augmented Reality results

In Augmented Reality, the main aspect analyzed for this work concerns the highlight of key buildings. During discussions, 44% of the subjects confirmed the interest of the highlights, independently of the test set they have experienced. Subject preferences are showed in figure 10.

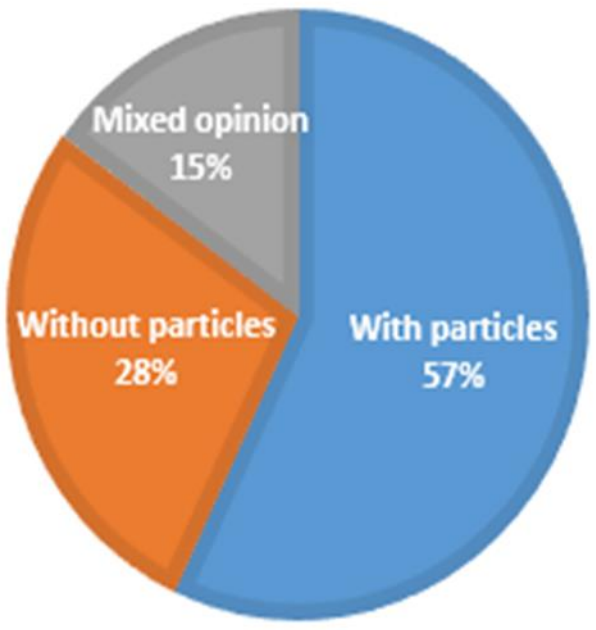

**Fig. 10.** Users preferences in Augmented Reality

Nevertheless, it seems important to underline that all the subjects having preferred the solution without highlights explained that it brought a visual overload. An animated solution, lighter and more discreet would probably be well accepted. We consider then than our third hypothesis can be validated.

### 4.4 PERSPECTIVES

The interest of a loading cursor has been confirmed in our study. However, it slightly lacks visibility; certain users needed a time before noticing it. It could be interesting to emphasize the loading cursor by a different color than the original cursor on which it is inserted.

As well in Virtual as in Augmented Reality, the necessity of highlighting the key buildings was underlined. By taking into account the results of the experiments, as well as the users remarks and suggestions, it could be interesting to position a ñrealò animated element floating on top every key building instead a visual effect. In the style of a sign on roof, this animated element would indicate the building by an icon.

These elements will allow us to improve the application interaction. Future work will allow users to be able to access to the information, in the form of web pages at the first place, when they interact with a key building.

## 5 CONCLUSIONS

This paper presents Immercity, a content curation application in Virtual and Augmented Reality allowing to structure and share information by using a 3D city.
Prototypes were developed in order to validate the visual cues and pointing implemented, which allows users to identify and move towards key buildings.
The results of our experiments provide insights about the interest and usage of the visual cues for selection and navigation tasks. Future research needs to confirm these results by studying the particular design characteristics which contribute to a more intuitive interaction with the city elements.